\documentclass[fleqn,usenatbib,useAMS]{mnras}

\usepackage{graphicx}	
\usepackage{amsmath}
\usepackage{amssymb}	
\usepackage{physics} 
\newcommand{\p}{\partial}
\newcommand{\OmK}{\Omega_\text{K}}

\newcommand{\nn}{\nonumber}
\newcommand\bb[1]{\mbox{\boldmath{$#1$}}}
\usepackage[T1]{fontenc}

\defcitealias{cui+24}{Paper I}

\begin{document}
%%%%%%%%%%%%%%%%%%% TITLE PAGE %%%%%%%%%%%%%%%%%%%
\title[]{Rossby wave instability in weakly ionized protoplanetary disks. II. radial B-fields}

\author[]{Can Cui$^{1}$\thanks{\href{mailto:can.cui@astro.utoronto.ca
}{can.cui@astro.utoronto.ca}} and Zijin Wang$^{2}$
\\ \\
$^{1}$Department of Astronomy and Astrophysics, University of Toronto, Toronto, ON M5S 3H4, Canada \\
$^{2}$Department of Mathematics, University of Toronto, Toronto, ON M5S 2E4, Canada \\
}

\pubyear{2024}

\label{firstpage}
\pagerange{\pageref{firstpage}--\pageref{lastpage}}
\maketitle

\begin{abstract}

In the first paper of this series, \citetalias{cui+24} investigated the influence of pure azimuthal and vertical magnetic fields on the linear growth of the Rossby Wave Instability (RWI). In this second paper, we extend the analysis to examine the effect of radial magnetic fields on RWI modes, incorporating all three non-ideal MHD effects - Ohmic resistivity, Hall drift, and ambipolar diffusion. The presence of a radial field relies on the disk's vertical shear and vertical magnetic field. We perform radially global linear analyses and solve the matrix eigenvalue problems using the spectral code \textsc{Dedalus}. It is found that in the ideal MHD limit, radial fields amplify the suppressive effect of vertical fields on RWI growth rates, with reductions occurring at relatively weak field strengths ($\beta \sim 10^3 - 10^4$), applicable to protoplanetary disks. In the non-ideal MHD regime, when any of the three non-ideal effects become sufficiently strong, the growth rates revert to their hydrodynamic values.

\end{abstract}


%%%%%%%%%%%%%%%%%%%%%%%%%%%%%%%%%%%%%%%%%%%%%%%%%%
\section{Introduction}\label{sec:in}

Protoplanetary disks are the birth place of planets, where dust particles coagulate and collapse into planetesimals. Planet formation involves the growth of micron-sized dust into km-sized planets \citep{armitage11,simon+22,Joanna+23}. In the early stage of planet formation, grains of $0.1-1\mu m$ in size grow through mutual collisions and sticking \citep{bw08,gb15,blum18}. This process leads to the formation of mm- to cm-sized grains, or pebbles \citep{bw08,gb15,blum18}. The intermediate stage involves the conversion of pebbles to km-sized planetesimals. However, the growth by sticking is halted by the bouncing or fragmentation \citep{guttler+10,zsom+10} in the inner disk, and by the fast radial drift in the outer disk \citep{weidenschilling77,birnstiel_etal10,birnstiel24}. 

The Rossby wave instability (RWI) may provide a promising solution to the dust growth barriers. Numerical simulations suggested that the RWI can generate large, crescent-shaped vortices in azimuth \citep{godon99,li+01,meheut+12}. These vortices concentrate dust grains towards their centers, where the pressure maxima reside. The concentration of dust can subsequently trigger streaming instability and gravitational collapse, facilitating the processes of planetesimal formation \citep{gw73,yg05}. The non-linear stage of RWI may account for the azimuthal asymmetries observed in (sub-)millimeter dust continuum and CO rotational transition lines by ALMA \citep{huang_etal18,vdmarel+21}.

To excite RWI modes, the necessary condition is the presence of local extrema in the disk's radial vortensity profile \citep{lovelace99,li_etal00,chang+23,cy24}. RWI drives the exponential growth of non-axisymmetric modes on each side of the corotation \citep{tl08}. The unstable Rossby modes are confined between the inner and outer Lindblad resonances, where density waves are launched \citep{shu+64,gt79}. In the context of protoplanetary disks, numerical simulations of RWI have been performed in various scenarios, including gap edges carved by a planet \citep{zhu+14,zs14,hammer+17,li+20,cimerman+23}, dead zone edges of the magneto-rotational instability \citep[MRI;][]{lyra12,miranda+16,miranda+17}, or more recently in magnetically induced rings and gaps \citep{Hsu+24}. These locations are where local vortensity extrema typically occur in the disk.

Large-scale, ordered magnetic fields are thought to thread protoplanetary disks, originating from primordial molecular clouds \citep{galli93, girart+06, girart+09}. These magnetic fields are crucial for disk evolution, driving angular momentum transport through mechanisms including magnetized winds, MRI turbulence, and laminar magnetic stresses \citep{bs13,lesur21}. In the ideal magnetohydrodynamics (MHD) regime, the magnetic fields are considered fully ionized and coupled perfectly to the gas. However, the coupling between gas and magnetic fields in protoplanetary disks is inefficient because of the weak thermal ionization by the irradiation of the central star, and the weak non-thermal ionization by the stellar FUV, EUV, X-rays and cosmic rays \citep{wardle07,bai11a}. As a result, ideal MHD is not applicable in the disk, and three non-ideal MHD effects must be invoked to accurately describe the gas dynamics -- Ohmic resistivity, Hall drift, and ambipolar diffusion \citep{lesur21}.

Ohmic resistivity arises from electron-neutral collisions, whereas the Hall drift and ambipolar diffusion result from the relative drift between electrons and ions and between neutrals and ions, respectively. If present, the magneto-rotational instability (MRI) can drive strong turbulence in the disk, potentially suppressing the RWI \citep{cb22}. Fortunately, non-ideal MHD effects can effectively dampen the MRI \citep{lk22}. In the inner disk, MRI is suppressed at the midplane by Ohmic resistivity and at the disk surface by ambipolar diffusion \citep[e.g.,][]{gressel_etal15}. In the outer disk, ambipolar diffusion dominates among non-ideal MHD effects, damping MRI at the midplane \citep[e.g.,][]{simon_etal13a,simon_etal13b}. The Hall drift further complicates the picture, as the non-linear evolution of the Hall-dominated disk depends on the alignment between the vertical magnetic fields and the disk rotation \citep[e.g.,][]{bethune+16,bethune_etal17,bai17}.

The RWI has predominantly been studied in the hydrodynamic regime. However, understanding how magnetic fields influence RWI modes is crucial. In the first paper of this series, \citetalias{cui+24} examined the RWI under constant azimuthal or vertical magnetic fields. Our findings showed that in the ideal MHD regime, strong magnetic fields suppress RWI growth rates, while non-ideal MHD effects can revive the instability. The influence of the Hall effect introduces additional complexity, as the sign of the Hall Els\"{a}sser number plays a role in modifying the results. In this work, we extend the previous study to incorporate radial magnetic fields, which naturally arise in the presence of both vertical shear and vertical magnetic fields.

This paper is organized as follows. In Section \ref{sec:th}, we introduce the governing equations, equilibrium state, and perturbation equations. Section \ref{sec:me} details the numerical methods used to solve the ordinary differential equations (ODEs) that describe the magnetized RWI. In Section \ref{sec:re}, we present the numerical results and analyze the RWI growth rates. Finally, Section \ref{sec:cd} summarizes our main findings and provides discussion.

\section{Theory}\label{sec:th}

In this section, we present the basic equations and equilibrium state. Specifically, they are dynamical equations (\S\ref{sec:de}), equilibrium solutions (\S\ref{sec:es}), and perturbation equations (\S\ref{sec:pe}).

\subsection{Dynamical equations}\label{sec:de}

We study the stability of a 3D, compressible, magnetized disk in cylindrical coordinates ($r, \phi, z$). Disk self-gravity is neglected. The gravitational potential is given by $\Phi=-GM_\star/(r^2+z^2)^{1/2}$, where $M_\star$ is the mass of the central star. Then, the dynamical equations of mass, momentum, and entropy conservation are
\begin{equation}
\frac{d\rho}{dt}+\rho\nabla\cdot\bb{v}=0,
\label{eq:1}
\end{equation}
\begin{equation}
\frac{d\bb{v}}{dt} + \frac{1}{\rho}\nabla\bigg[P+\frac{B^2}{8\pi}\bigg]+\nabla\Phi - \frac{1}{4\pi\rho}(\bb{B}\cdot\nabla)\bb{B}=0,
\label{eq:2}
\end{equation}
\begin{equation}
\frac{dS}{dt}=0.
\label{eq:3}
\end{equation}
The induction equation in Gaussian units is 
\begin{equation}
\frac{\p\bb{B}}{\p t}-\nabla\times(\bb{v}\times\bb{B}-c\bb{E}')=0.
\label{eq:ind}
\end{equation}
The material derivative is defined as $d/dt\equiv\p/\p t+v\cdot\nabla$, and $S\equiv P/\rho^\Gamma$ is the entropy of the disk matter.

The components of non-ideal MHD terms manifest in the induction Equation \eqref{eq:ind}. The electric field in the rest fluid frame is 
\begin{equation}
\bb{E}^\prime=\frac{4\pi}{c^2}{(\eta_O\vb{J} +\eta_\mathrm{H}\bb{J}\times\bb{b}+\eta_A\vb{J_\perp})}\ ,
\end{equation}
where Ohmic, Hall and ambipolar diffusivities are denoted by $\eta_O$, $\eta_H$ and $\eta_A$. The current density is given by $\vb{J}=c\curl\vb{B}/4\pi$, where we express the component of $\vb{J}$ that is perpendicular to the magnetic field by $\vb{J_\perp} = - (\vb{J} \times \bb{b}) \times \bb{b}$, and the unit vector of magnetic field is denoted by $\bb{b}=\vb{B}/B$. 

The diffusivities in terms of conductivities are written as 
\citep{wardle07,wang_etal19,lesur21}
\begin{equation}
\eta_\mathrm{O} = \frac{c^2}{4\pi}\left(\frac{1}{\sigma_\mathrm{O}}\right),
\end{equation}   
\begin{equation}
\eta_\mathrm{H} = \frac{c^2}{4\pi}\left(\frac{\sigma_\mathrm{H}}{\sigma_\mathrm{H}^2+\sigma_\mathrm{P}^2}\right),
\end{equation}
\begin{equation}
\eta_\mathrm{A} = \frac{c^2}{4\pi}\left(\dfrac{\sigma_\mathrm{P}}{\sigma_\mathrm{H}^2+\sigma_\mathrm{P}^2}\right) - \eta_\mathrm{O},
\end{equation}
where $\sigma_\mathrm{O}$, $\sigma_\mathrm{H}$ and $\sigma_\mathrm{P}$ are Ohmic, Hall, and Pederson conductivities. For $j^\mathrm{th}$ charged species, let $Z_je$ being the charge and $n_j$ the number density, then
\begin{equation}
\sigma_\mathrm{O} = \frac{ec}{B} \sum_j n_jZ_j\beta_j,
\end{equation}
\begin{equation}
\sigma_\mathrm{H} = \frac{ec}{B} \sum_j \frac{n_jZ_j}{1+\beta_j^2},
\end{equation}
\begin{equation}
\sigma_\mathrm{P} = \frac{ec}{B} \sum_j\frac{n_jZ_j\beta_j}{1+\beta_j^2},
\end{equation}
in which $\beta_j$ is the Hall parameter defined as the ratio of the gyrofrequency to the collision rate with neutrals,
\begin{equation}
\beta_j = \dfrac{Z_jeB}{m_jc}\frac{1}{\gamma_j\rho},
\end{equation}
and 
\begin{equation}
\gamma_j=\frac{\langle\sigma v\rangle_j}{m_n + m_j},
\end{equation}
where $m_j$ is the molecular mass of charged species, $m_n$ is the mean molecular mass of the neutrals, and $\langle\sigma v\rangle_j$ is the momentum exchange rate between the $j^\mathrm{th}$ species and the neutrals.

Finally, we introduce the dimensionless Els\"{a}sser numbers that quantify the strengths of non-ideal MHD effects \citep{cb20,cb21},
\begin{equation}
\Lambda=\frac{v_\textrm{A}^2}{\eta_\mathrm{O}\OmK}, \qquad 
\mathrm{Ha}=\frac{v_\mathrm{A}^2}{\eta_\mathrm{H}\OmK}, \qquad 
{\rm Am}=\frac{v_\mathrm{A}^2}{\eta_\mathrm{A}\OmK},
\end{equation}
where the Alfv\'{e}n velocity is $v^2_{\mathrm A}=B^2/4\pi\rho$, and $\OmK$ is the Keplerian angular speed. Note that the Els\"{a}sser numbers are inverse proportional to diffusivities, and that $\Lambda$ and Ha are $B$-dependent, as $\eta_\mathrm{O}\propto \mathrm{const}$, $\eta_\mathrm{H}\propto B$, $\eta_\mathrm{A}\propto B^2$. 

\subsection{Equilibrium state}\label{sec:es}

The equilibrium disk model is stationary ($\p/\p t=0$) and axisymmetric ($\p/\p \phi=0$). The model is radially global and vertically local, and hence all background quantities are independent of $z$. 
%The presence of radial magnetic fields require $z\neq 0$, where the vertical shear vanishes \citep{bl15}. 
The physical quantities in the equilibrium state are denoted by subscript ``0''. The velocity field in equilibrium has only the azimuthal component $\bb{v_0}=(0,v_{\phi0},0)$. The magnetic field $\bb{B}_0=(B_{r0},0,B_{z0})$ is taken to be a constant. The unit vector of the background magnetic field is then $\bb{b_0}=(b_{r0},0,b_{z0})=(B_{r0}/|B|,0,B_{z0}/|B|)$. Achieving magnetic equilibrium in cylindrical coordinates can be complicated, because of the presence of curvature terms. Here, we opt for neglecting the curvature terms for the steady state.

We first establish vortensity extrema, which is the necessary condition to excite the RWI \citep{cy24}. We follow \citetalias{cui+24} and introduce a Gaussian bump centered at $r = r_0$ in the density profile,
\begin{equation}
\frac{\rho_\mathrm{0}}{\rho_{00}} = 1+ (A-1)\exp\bigg[-\frac{1}{2}\bigg(\frac{r-r_0}{\Delta r}\bigg)^2\bigg],
\end{equation}
where $\rho_{00}$ represents the background density profile without the Gaussian bump and is assumed to be constant for simplicity. We assume a barotropic flow, where pressure is related to density by
\begin{equation}
\frac{P_0}{P_{0\ast}}= \bigg[\frac{\rho_0}{\rho_{0\ast}}\bigg]^\Gamma,
\end{equation}
and subscript ``$0\ast$'' denotes background quantities evaluated at $r_0$. The adiabatic index is denoted by  $\Gamma$, and the adiabatic sound speed is defined as $c_{s0}\equiv(\Gamma P_0/\rho_0)^{1/2}$. 
By specifying the disk aspect ratio to be $c_{s0\ast}/v_\mathrm{K0\ast}=0.06$, we can obtain $P_{0\ast}$, and subsequently $P_0$. We follow \citetalias{cui+24} and set $GM = \rho_0 = r_0=1$, and the parameters are $\Delta r/r_0=0.05$, $\Gamma=5/3$, and $A=1.5$. 
The azimuthal velocity is computed from the radial momentum equation 
\begin{equation}
\frac{v_{\phi 0}^2}{r}=\frac{1}{\rho_0}\frac{\p P_0}{\p r} + \frac{\p\Phi}{\p r}.
\end{equation}

By neglecting curvature terms, the steady-state solution allows for a uniform background magnetic field. This equilibrium solution is analogous to that in the shearing box model of \citet{lp18}. We describe this equilibrium in detail below, focusing first on the ideal MHD regime. Since the curvature terms are omitted, the divergence-free condition $\nabla \cdot \bb{B} = 0$ is easily satisfied. Then, the components of the magnetic field should satisfy 
\begin{equation}
\bb{B_0}\cdot\nabla \bb{v_0}=0,
\end{equation}
which is obtained from the y-component of the induction equation. This gives
\begin{equation}
\frac{B_{r0}}{B_{z0}} = - \frac{q_z}{q_r},
\label{eq:12}
\end{equation}
where the shears of the angular speed are
\begin{equation}
q_r \equiv -\frac{\p\ln\Omega}{\p\ln r},
\end{equation}
\begin{equation}
q_z \equiv - \frac{r\p\ln\Omega}{\p z}.
\end{equation}

Eq. \eqref{eq:12} is equivalent to eq. (10) in \citet{cui+24} by eliminating the curvature terms. The dimensionless radial shear is parameterized by $q_r$. It is $q_r=3/2$ for a Keplerian disk. The presence of $B_{r0}$ relies on the vertical shear of the disk, parameterized by the dimensionless quantity $q_z$. Unlike $q_r$, which is generally positive for protoplanetary disks, the sign of $q_z$ reverses with respect to the midplane \citep{bl15}. Above the midplane $z>0$, $q_z > 0$, and below the midplane $z<0$, $q_z < 0$. In protoplanetary disks, $|q_z| \sim h$, leading to $|B_{r0}/B_{z0}| \ll 1$. In magnetic equilibrium, we set $q_r = 3/2$ and $|q_z| = h$ for simplicity. When seeking numerical solutions, we set $B_{z0}$ to a constant, and $B_{z0} > 0$ without loss of generality. The strength of $B_{z0}$ is parameterized by the vertical plasma $\beta$, defined as the ratio of gas pressure to vertical magnetic pressure, $\beta=8\pi P_0/B_{z0}^2$. Given constant background magnetic fields, the current density $\vb{J}$ vanishes. Hence, the equilibrium in the non-ideal limit is easily obtained (see eq. \eqref{eq:ind}).

\subsubsection{Solberg-Hoiland criterion}\label{sec:}

The stability of the equilibrium disk, including the Gaussian bump, against axisymmetric adiabatic perturbations shall be examined. For hydrodynamic flow, the general condition for stability is given by the Solberg-Hoiland criterion. The stability is ensured if \citep{tassoul78}
\begin{equation}
\kappa^2+N^2 \geq 0,
\end{equation}
where the epicyclic frequency squared is defined as 
\begin{equation}
\kappa^2 = \frac{1}{r^3}\frac{dr^4\Omega^2}{dr},
\end{equation}
and $N$ is the buoyancy frequency,
\begin{equation}
N^2 = N_r^2 + N_z^2, 
\end{equation}
in which 
\begin{equation}
N_r^2 = -\frac{1}{\Gamma\rho}\frac{\p P}{\p r}\frac{\p\ln S}{\p r}, \qquad 
N_z^2 = -\frac{1}{\Gamma\rho}\frac{\p P}{\p z}\frac{\p\ln S}{\p z}.
\end{equation}
For a barotropic flow as employed in this work, $N^2$ is zero, and the condition becomes $\kappa^2\geq 0$, which is satisfied for the set of parameters chosen in this work.

\subsection{Perturbation equations}\label{sec:pe}

Consider small perturbations to eqs. \eqref{eq:1}-\eqref{eq:ind}, such that the physical quantities are composed of a background state and a perturbation part. For example, $\bb{v} = \bb{v}_0+\delta\bb{v}(r,z,\phi,t)$ and $P = P_0+\delta P(r,z,\phi,t)$. We consider Eulerian perturbations of the form $\propto f(r)\exp(ik_zz+im\phi-i\omega t)$, where $k_z$ is the vertical wavenumber, $m$ is the azimuthal mode number, and $\omega=\omega_r+i\gamma$ is the mode frequency, with $\gamma$ denoting the growth rate. We further define the Doppler-shifted wave frequency $\Delta\omega=\omega-m\Omega$, and the azimuthal wavenumber $k_\phi=m/r$. 

We now drop subscript ``$0$'' for background quantities throughout the rest of the paper. Our model encompasses eight perturbed quantities: $\delta\bb{v},\delta\bb{B},\delta\rho,\delta\Psi$. The perturbed enthalpy is defined as
\begin{equation}
\delta\Psi = \frac{\delta P}{\rho},
\label{eq:psi}
\end{equation}
and its derivative is
\begin{equation}
\frac{\p\delta\Psi}{\p r} = \frac{1}{\rho}\frac{\p\delta P}{\p r} - \frac{1}{\rho}\frac{\p\rho}{\p r}\delta\Psi.
\end{equation}
Furthermore, the length scales of entropy, pressure, and density variations are 
\begin{equation}
L_S \equiv \frac{\Gamma}{d \ln S/dr}, \qquad
L_P \equiv \frac{\Gamma}{d \ln P/dr}, \qquad
L_\rho \equiv \frac{1}{d \ln\rho/dr}.
\end{equation}
These length scales are simply related by
\begin{equation}
\mathrm{\frac{1}{L_P} = \frac{1}{L_S} + \frac{1}{L_\rho}}.
\end{equation}
For a barotropic flow, the length scale of entropy approaches infinity, $1/\mathrm{L_S}\rightarrow 0$.

It follows that the perturbed continuity equation is written as 
\begin{equation}
\frac{\p\delta v_r}{\p r} + \bigg[\frac{1}{r}+\frac{1}{\mathrm{L_\rho}}\bigg]\delta v_r + ik_\phi\delta v_\phi - i\Delta\omega \frac{\delta\Psi}{c_s^2} = 0.
\end{equation}
The $r,\phi,z$-components of the momentum equation are 
\begin{align}
-i\Delta\omega&\delta v_r - 2\Omega\delta v_\phi + \frac{\p\delta\Psi}{\p r} +\frac{B_z}{4\pi\rho}\bigg[ \frac{\p\delta B_z}{\p r}-ik_z\delta B_r \bigg] = 0,
\end{align}
\begin{align}
-i\Delta\omega\delta v_\phi & + \frac{\kappa^2}{2\Omega}\delta v_r + ik_\phi\delta \Psi \nn \\  
& + \frac{1}{4\pi\rho}\bigg[ik_\phi(B_r\delta B_r+B_z\delta B_z) - B_r\bigg(\frac{\p\delta B_\phi}{\p r} + \frac{\delta B_\phi}{r}\bigg) \nn \\   
& - ik_zB_z\delta B_\phi \bigg] = 0,
\end{align}
\begin{align}
-i\Delta\omega\delta v_z + ik_z\delta\Psi + \frac{B_r}{4\pi\rho}\bigg[ ik_z\delta B_r - \frac{\p \delta B_z}{\p r} \bigg] = 0,
\end{align}
The entropy equation reads 
\begin{equation}
\delta \Psi = c_s^2\frac{\delta\rho}{\rho}.
\label{eq:25}
\end{equation}

The three-components of the induction equation are 
\begin{equation}
-i\Delta\omega\delta B_r - ik_z B_z\delta v_r + \bigg[ \frac{\delta v_r}{r} + ik_\phi\delta v_\phi + ik_z \delta v_z \bigg] B_r + C_1 = 0,
\end{equation}
\begin{align}
-i\Delta\omega\delta B_\phi & +  \bigg[ \frac{\delta v_\phi}{r} - \frac{\p\delta v_\phi}{\p r} \bigg] B_r + \bigg[ \frac{v_\phi}{r} - \frac{\p v_\phi}{\p r} \bigg] \delta B_r  \nn \\
& - ik_zB_z\delta v_\phi + C_2 = 0,
\end{align}
\begin{equation}
-i\Delta\omega\delta B_z -  \frac{\p\delta v_z}{\p r} B_r + \bigg[ \frac{\p\delta v_r}{\p r} + \frac{\delta v_r}{r} + ik_\phi\delta v_\phi \bigg] B_z + C_3 = 0,
\end{equation}
where $C_1, C_2, C_3$ involve the non-ideal MHD terms that will be expressed in the next subsection.
 
\subsubsection{non-ideal MHD limit}

We now derive the perturbation equations for non-ideal MHD effects, 
\begin{align}
C_1 = & - \eta_\mathrm{O}\bigg[\frac{\p^2\delta B_r}{\p r^2} - (k_\phi^2+k_z^2)\delta B_r \bigg] \nn \\ 
& +\eta_\mathrm{H} \bigg[ib_zk_z + b_r\frac{\p}{\p r}\bigg] [ik_\phi\delta B_z - ik_z\delta B_\phi] \nn \\ 
& - \eta_\mathrm{A} \bigg[\bigg(-b_z^2k_z^2 + b_r^2\frac{\p^2}{\p r^2} + 2b_rb_zik_z\frac{\p}{\p r}\bigg) \delta B_r \nn \\ 
& + \bigg(\frac{\p^2}{\p r^2} - k_\phi^2 - k_z^2\bigg)(\delta B_rb_r + \delta B_zb_z) b_r\bigg],  \nn \\ 
\label{eq:32}
\end{align}
\begin{align}
C_2 = & - \eta_\mathrm{O} \bigg[\frac{\p^2\delta B_\phi}{\p r^2} - (k_\phi^2+k_z^2)\delta B_\phi \bigg] \nn \\ 
& +\eta_\mathrm{H} \bigg[ib_zk_z + b_r\frac{\p}{\p r}\bigg] \bigg[ik_z\delta B_r - \frac{\p \delta B_z}{\p r}\bigg] \nn \\ 
& - \eta_\mathrm{A} \bigg[-b_z^2k_z^2 + b_r^2\frac{\p^2}{\p r^2} + 2b_rb_zik_z\frac{\p}{\p r}\bigg]\delta B_\phi,  \nn \\ 
\end{align}
\begin{align}
C_3 = & - \eta_\mathrm{O}\bigg[\frac{\p^2\delta B_z}{\p r^2} - (k_\phi^2+k_z^2)\delta B_z\bigg] \nn \\ 
& +\eta_\mathrm{H}\bigg[ib_zk_z + b_r\frac{\p}{\p r}\bigg] \bigg[\frac{\p\delta B_\phi}{\p r} - ik_\phi\delta B_r \bigg] \nn \\ 
& - \eta_\mathrm{A} \bigg[\bigg(-b_z^2k_z^2 + b_r^2\frac{\p^2}{\p r^2} + 2b_rb_zik_z\frac{\p}{\p r}\bigg) \delta B_z \nn \\ 
& + \bigg(\frac{\p^2}{\p r^2} - k_\phi^2 - k_z^2\bigg)(\delta B_rb_r + \delta B_zb_z) b_z\bigg].  \nn \\ 
\label{eq:34}
\end{align}
We omit the curvature terms in $C_1$, $C_2$, and $C_3$ for simplicity. As shown in Appendix \ref{app:c}, curvature terms only slightly affect the growth rates. Compared to a disk in equilibrium with a purely vertical magnetic field \citepalias{cui+24}, Ohmic resistivity possesses the same non-ideal MHD terms with the presence of radial magnetic field. However, the inclusion of a radial field introduces additional components to the Hall and ambipolar diffusion terms.

\section{Numerics}\label{sec:me}

We solve the linearized equations numerically using pseudospectral methods, described in \citetalias{cui+24}. Pseudospectral method approximates solutions to differential equations by a weighted sum of orthogonal basis functions \citep{boyd}. The differential equations in \S\ref{sec:pe} formulate standard linear eigenvalue problems (EVPs). They can be expressed compactly in a generalized matrix form,
\begin{equation}
A\vec{x} = L\vec{x} + \omega M\vec{x} = 0,
\end{equation} 
where $\omega$ represents the eigenvalue, $\vec{x} = [\vec{\delta v}_r, \vec{\delta B}_r, \vec{\delta \rho}, \ldots]^\mathrm{T}$ is a vector of eigenfunctions containing $8$ perturbed quantities, and $A$, $L$, and $M$ are $8N \times 8N$ matrices, with $L$ comprising linear operators.

We use \textsc{Dedalus}\footnote{\url{https://dedalus-project.org/}}, a general-purpose spectral code, to solve the linear EVPs \citep{burns+20}. We choose Chebyshev polynomials of the first kind, $T_n$, where $n = 0, 1, 2, \ldots, N-1$, as the orthogonal basis. The radial domain, spanning $r \in [0.4, 1.6]$, is discretized into $N$ collocation points. These non-uniform nodes are computed by the roots of the $N$th-degree Chebyshev polynomial, $T_N$. We employ the dense solver method via \texttt{solve\_dense} in \textsc{Dedalus}, which converts matrix $A$ into dense arrays, and utilize \texttt{scipy.linalg.eig} routine from Python, that can solve EVPs directly. We employ a numerical resolution of $N = 256$, and increase the resolution by a factor of 1.5 in order to filter out the non-physical modes. 

\section{Results}\label{sec:re}

\begin{figure}
    \centering
    \includegraphics[width=0.5\textwidth]{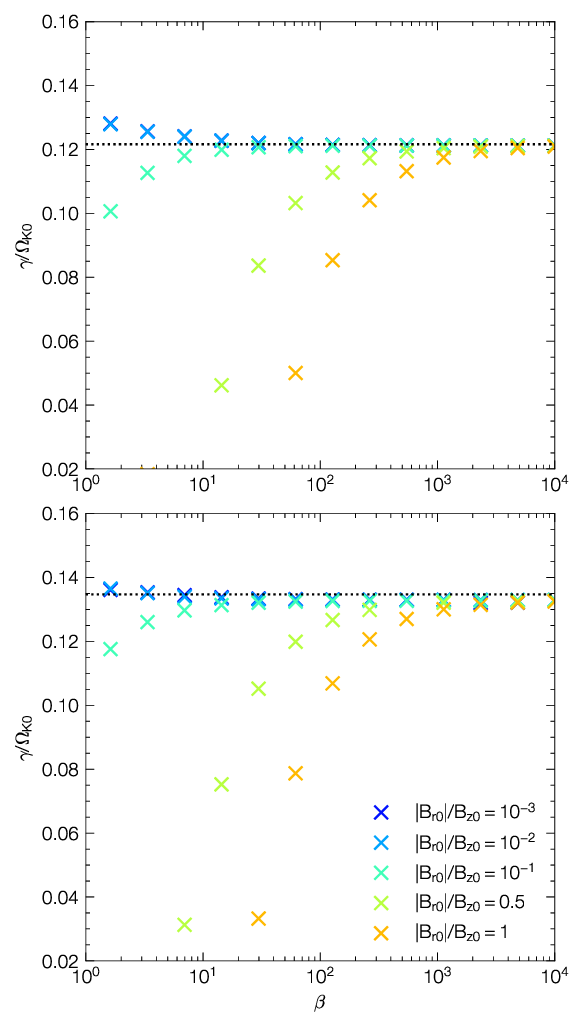}
    \caption{Normalized growth rate $\gamma/\Omega_\mathrm{K0}$ versus plasma $\beta$ in the ideal MHD limit for azimuthal mode numbers $m=3$ (top) and $m=4$ (bottom). Colors denote the ratio of radial to vertical magnetic field, 
    $|B_{r0}|/B_{z0}\in\{10^{-3},10^{-2},10^{-1},0.5, 1\}$. Dotted horizontal lines denote the hydrodynamic growth rates.}
    \label{fig:i}    
\end{figure}

\subsection{ideal MHD limit}

We begin with the ideal MHD regime and investigate the effect of radial magnetic fields on RWI growth rates. Figure \ref{fig:i} shows growth rates as a function of plasma $\beta$. We fix the azimuthal mode number at $m=3$ (top panel) and $m=4$ (bottom panel). Dotted horizontal lines denote the growth rates in the hydrodynamic limit. Recall that large (small) $\beta$ corresponds to weak (strong) magnetic fields. 

It is clearly seen that in both panels, when the radial magnetic field is relatively weak compared to the vertical component, $|B_{r0}|/B_{z0}\lesssim 10^{-2}$, the growth rates closely follow those of the pure vertical magnetic field model (pure $B_z$ model) in \citetalias{cui+24}. However, for stronger radial fields, with $|B_{r0}|/B_{z0} \gtrsim 10^{-1}$, the growth rates decline. This decrease is more significant at lower $\beta$. The physical mechanism would be, in the ideal MHD limit, gas and magnetic fields are perfectly coupled. The magnetic tension impedes the rolling up of vortex sheets.
Our result is in contrast with \citet{yu13}, who found the growth rates increase with decreasing $\beta$ (see their Figure 2 and Figure 5). The origin of this discrepancy remains unclear.

In \citetalias{cui+24}, we explored two disk models: the pure $B_z$ model and the pure $B_\phi$ model. In the former, the growth rates remain largely unchanged for $\beta > 1$, whereas in the latter, they begin to decline at weaker field strengths, around $\beta \sim 100$. Figure \ref{fig:i} indicates that the presence of a radial magnetic field further dampens the growth rates in the $B_z$ model. Global MHD simulations suggest that radial and vertical field components reach comparable strengths in quasi-steady states \citep[e.g.,][]{bethune_etal17,gressel_etal20,cb21,cb22}. When $|B_{r0}| \approx |B_{z0}|$ (orange crosses in Figure \ref{fig:i}), the growth rates start to decrease from their hydrodynamic values at $\beta \sim 10^4$, which is the typical field strength in protoplanetary disks \citep{lesur21}. Therefore, for real disks, particularly those with strong magnetic fields, RWI modes might be weakened in regions where ideal MHD dominates.

\begin{figure}
    \centering
    \includegraphics[width=0.5\textwidth]{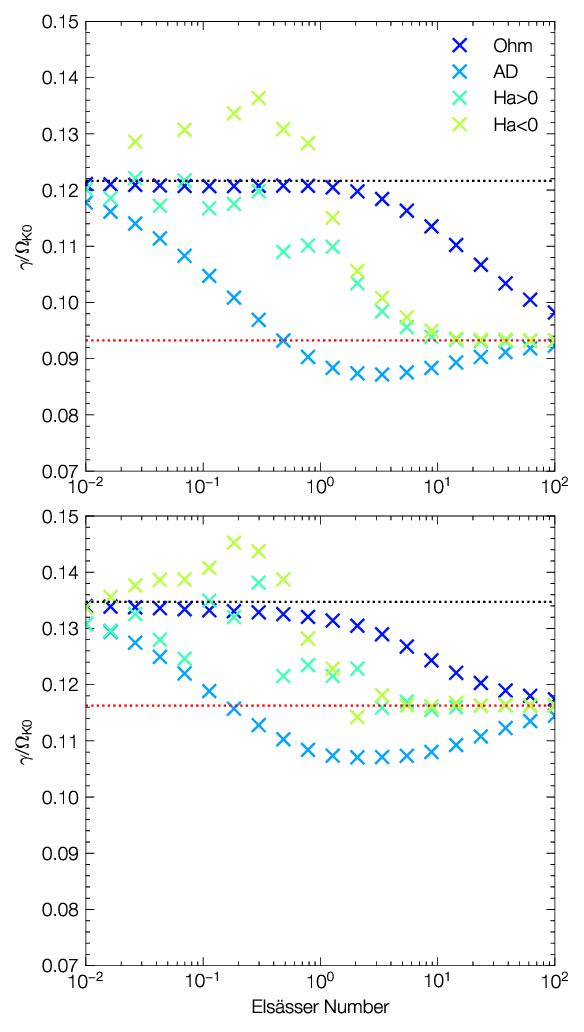}
    \caption{Normalized growth rate versus Els\"{a}sser number at $\beta=1$ and $|B_{r0}|/B_{z0}=0.1$ in the non-ideal MHD limit. Azimuthal mode number is fixed at $m=3$ (top) and $m=4$ (bottom). Dotted horizontal lines denote the hydrodynamic (black) and ideal MHD (red) growth rates.
    }
    \label{fig:n1}    
\end{figure}

\begin{figure}
    \centering
    \includegraphics[width=0.5\textwidth]{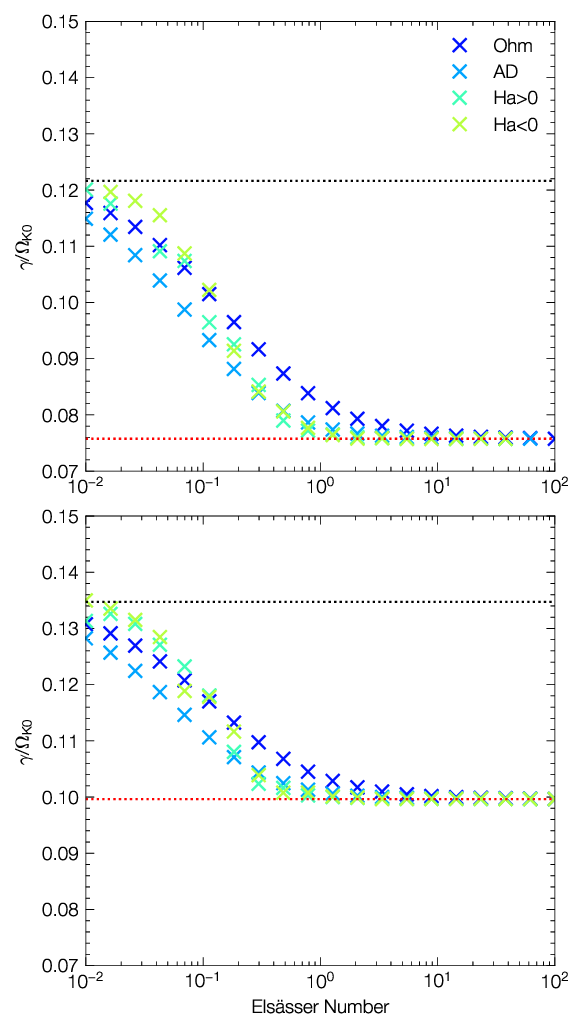}
    \caption{Same as Figure \ref{fig:n1} but for $\beta=100$ and $|B_{r0}|/B_{z0}=1$.}
    \label{fig:n2}    
\end{figure}

\subsection{non-ideal MHD limit}

In \citetalias{cui+24}, we found that in the strong non-ideal MHD regime, characterized by small Els\"{a}sser numbers, the growth rates recover the hydrodynamic results. Here, we examine how non-ideal MHD effects influence the growth rates in the presence of radial magnetic fields. Figure \ref{fig:n1} shows the RWI growth rate as a function of Els\"{a}sser number for $\beta=1$ and $|B_{r0}|/B_{z0}=0.1$. The top panel corresponds to an azimuthal mode number of $m=3$, while the bottom panel corresponds to $m=4$. Figure \ref{fig:n2} presents a similar plot but for $\beta=100$ and $|B_{r0}|/B_{z0}=1$.

Both Figures illustrate that as the Els\"{a}sser number approaches $\sim 100$, the growth rates converge to those in the ideal MHD limit (dotted red lines). This occurs because large Els\"{a}sser number indicates weak non-ideal MHD, allowing the system to approach the ideal MHD regime. Conversely, when the Els\"{a}sser number decreases to $\sim 0.01$, the RWI modes closely resemble the hydrodynamic results (dotted black lines). This can be attributed to the fact that strong non-ideal effects weaken the coupling between gas and magnetic fields, allowing the gas to move freely rather than strictly following the magnetic field lines.

For intermediate Els\"{a}sser numbers ($0.01 \lesssim \mathrm{Els} \lesssim 100$), the growth rates generally transition between the hydrodynamic and ideal MHD limits, as shown in both Figures. However, in Figure \ref{fig:n1}, we observe that in the ambipolar diffusion-dominated disk, the growth rates can fall below the ideal MHD limit (dotted red line). Additionally, for cases with $\mathrm{Ha}<0$, the growth rates can exceed the hydrodynamic limit (dotted black line). 
To further explore this behavior, we plotted the growth rates for $\beta = 10$ and $|B_{r0}|/B_{z0} = 0.5$, a parameter choice lies in between those used in Figures 2 and 3. We find that at higher $\beta$, the transition between the hydrodynamic and ideal MHD regimes becomes smoother for the Hall effect, and for ambipolar diffusion, the growth rates tend to remain bounded between the two regimes. The underlying reason for this behavior, however, remains unclear.

\section{Conclusions and Discussion}\label{sec:cd}

In the first paper of this series, \citetalias{cui+24} investigated the effect of pure azimuthal ($B_\phi$) and vertical ($B_z$) magnetic fields on the RWI. In this study, we extend our analysis by incorporating radial magnetic fields into the background state and examining their impact on RWI linear modes using Eulerian perturbations. We consider both the ideal and non-ideal MHD regimes, accounting for the effects of Ohmic resistivity, Hall drift, and ambipolar diffusion. To solve the matrix eigenvalue problems, we employ spectral code \textsc{Dedalus}. 

It is found that in the ideal MHD limit, radial fields enhance the suppressive effect of vertical fields on RWI growth rates. This decrease in growth rates starts even at relatively weak field strengths, around $\beta \sim 10^3 - 10^4$, applicable to protoplanetary disks. In the non-ideal MHD regime, all three non-ideal effects, when sufficiently strong, restore growth rates to values comparable to hydrodynamic models.

Non-ideal MHD effects are expected to be significant across a large portion of the disk, particularly at radii beyond $\sim 0.1$ AU \citep{desch+15}. Our results therefore suggest that the RWI may emerge across a large part of the disk wherever vortensity extrema occur. Additionally, the vertical shear instability is a promising mechanism for driving turbulence in protoplanetary disks \citep{nelson_etal13,ly15,cl22,svanberg+22,dang+24}. Both the vertical shear instability and the streaming instability may develop under the non-ideal MHD regime \citep{cb20,cl21,xb22}, potentially coexisting with the RWI. Future numerical simulations may provide insight into the interplay between these instabilities and the nature of turbulence in protoplanetary disks.

Annular substructures are prevalent in protoplanetary disks \citep{huang_etal18,long_etal18}. They serve as ideal sites for planetesimal formation. Ring locations are particularly favorable for the onset of the Rossby wave instability (RWI) because it naturally gives rise to pressure bumps. However, ALMA observations indicated that only about 10\% of disks possessing rings exhibit crescent-like azimuthal asymmetries \citep{huang_etal18}. Since these observations trace continuum dust emission, it is important to consider the influence of dust grains on the development of the RWI \citep{lb23}. Future studies should explore this aspect by incorporating realistic pressure bump properties into linear analyses to assess whether the RWI can still operate in real disks.

One limitation of this work is the omission of curvature terms when deriving the equilibrium solutions. Future studies could refine this by obtaining steady-state solutions numerically, rather than the analytical approximation used here.
Another simplification is the assumption of a vertical wavenumber of zero ($k_z = 0$) throughout the linear analysis. A more accurate treatment would account for the disk's vertical structure, considering background quantities as a function of height. However, since the RWI is fundamentally a radially global problem, incorporating vertical stratification would lead to 2D matrix eigenvalue problems, which are prohibitively time-consuming to solve.

%%%%%%%%%%%%%%%%% Acknowledgements %%%%%%%%%%%%%%%%%%%%%

\section*{Acknowledgements}

CC acknowledges funding from Natural Sciences and Engineering Research Council of Canada. 

\section*{Data Availability}

The data underlying this article will be shared on reasonable request to the corresponding author.

%%%%%%%%%%%%%%%%% APPENDICES %%%%%%%%%%%%%%%%%%%%%
\appendix

\section{Curvature terms in linearized induction equations}\label{app:c}

\begin{figure}
    \centering
    \includegraphics[width=0.5\textwidth]{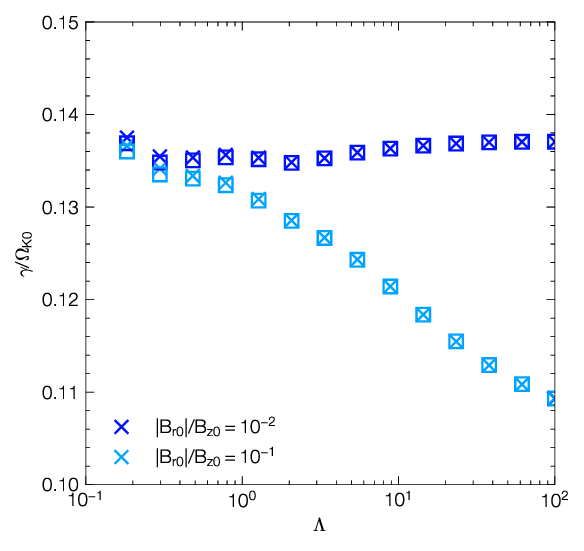}
    \caption{
    Growth rate versus Ohmic Els\"{a}sser number $\Lambda$ at $\beta=1$ and $|B_{r0}|/B_{z0}=0.01$ (dark blue) or $|B_{r0}|/B_{z0}=0.1$ (blue), with (squares) and without (crosses) curvature terms. 
    }
    \label{fig:c}    
\end{figure}

Figure \ref{fig:c} shows the growth rate versus Ohmic Els\"{a}sser number $\Lambda$ at $\beta=1$ for $|B_{r0}|/B_{z0}=0.01$ (dark blue) and $|B_{r0}|/B_{z0}=0.1$ (blue) with (squares) and without (crosses) curvature terms. It is clearly seen that the growth rates are largely unaffected when we include the curvature terms in the perturbed induction equations.

%%%%%%%%%%%%%%%%%%%% REFERENCES %%%%%%%%%%%%%%%%%%

\bibliographystyle{mnras}
\bibliography{disk} 

%%%%%%%%%%%%%%%%%%%%%%%%%%%%%%%%%%%%%%%%%%%%%%%%%%
\bsp	
\label{lastpage}
\end{document}